\documentclass[a4paper, conference]{IEEEtran}

\usepackage{amsmath,amsfonts}
\usepackage{algorithmic}
\usepackage{graphicx}
\usepackage{textcomp}
\usepackage{xcolor}
\usepackage{cleveref}
\usepackage{enumitem}
\usepackage[acronym]{glossaries}
\usepackage{multirow}
\usepackage{xspace}
\usepackage{microtype}
\usepackage{pifont}
\usepackage{caption}
\usepackage{booktabs}
\usepackage{changepage}
\usepackage{multicol}

\usepackage[free-standing-units,per-mode=repeated-symbol]{siunitx}
\usepackage[allow-number-unit-breaks]{siunitx}
\DeclareSIUnit\flop{FLOP}
\DeclareSIUnit\flops{FLOPS}
\DeclareSIUnit\gate{GE}
\DeclareSIUnit\op{OP}
\DeclareSIUnit\macu{MACU}
\DeclareSIUnit\ops{OPS}
\DeclareSIUnit\core{core}
\DeclareSIUnit\request{request}
\DeclareSIUnit\cycle{cycle}
\DeclareSIUnit\teraops{TOPS}
\DeclareSIUnit\ghz{GHz}
\DeclareSIUnit\mhz{MHz}
\DeclareSIUnit[number-unit-product = ]\percent{\%}

\newcommand\terapool[1]{\ensuremath{\text{TeraPool-SDR}_{#1}}}

\makeatletter \newcommand{\AddSpaceIfAnonymous}{\@ifclasswith{acmart}{anonymous}{\vspace{10mm}}{}} \makeatother

\definecolor{MidnightBlue}{HTML}{191970}
\definecolor{Mint}{HTML}{3EB889}
\definecolor{EnglishRed}{HTML}{A4515C}
\definecolor{SelectiveYellow}{HTML}{FFBA08}
\definecolor{CyanProcess}{HTML}{08B2E3}
\definecolor{OliveDrab7}{HTML}{4D4730}
\definecolor{Red}{HTML}{FF0000}
\colorlet{color1}{MidnightBlue}
\colorlet{color2}{Mint}
\colorlet{color3}{EnglishRed}
\colorlet{color4}{SelectiveYellow}
\colorlet{color5}{CyanProcess}
\colorlet{color6}{OliveDrab7}
\colorlet{colorAlert}{Red}

\newacronym{pe}{PE}{Processing Element}
\newacronym{ai}{AI}{Artificial Intelligence}
\newacronym{ml}{ML}{Machine Learning}
\newacronym{cpu}{CPU}{Central Processing Unit}
\newacronym{asic}{ASIC}{Application Specific Integrated Circuit}
\newacronym[longplural={Systems-on-Chip}]{soc}{SoC}{System-on-Chip}
\newacronym{fpga}{FPGA}{Field Programmable Gate Array}
\newacronym{asip}{ASIP}{Application Specific Instruction Processor}
\newacronym{gpp}{GPP}{General Purpose Processor}
\newacronym{gp}{GP}{general-purpose}
\newacronym{gpgpu}{GP-GPU}{General Purpose Graphics Processing Unit}
\newacronym{gpu}{GPU}{Graphics Processing Unit}
\newacronym{sm}{SM}{Streaming Multiprocessor}
\newacronym{cuda}{CUDA}{Compute Unified Device Architecture}
\newacronym{mpi}{MPI}{Message Passing Interface}
\newacronym{cots}{COTS}{Commercial-Off-The-Shelf}
\newacronym{soa}{SoA}{state-of-the-art}
\newacronym{roi}{ROI}{Return on Investments}

\newacronym{lte}{LTE}{Long Term Evolution}
\newacronym{nr}{NR}{New Radio}
\newacronym{4g}{4G}{4th Generation}
\newacronym{5g}{5G}{5th Generation}
\newacronym{6g}{6G}{6th Generation}
\newacronym{3gpp}{3GPP}{3rd Generation Partnership Project}
\newacronym{oran}{O-RAN}{Open Radio Access Networks}
\newacronym{ran}{RAN}{Radio Access Networks}
\newacronym{cran}{C-RAN}{Cloud Radio-Access-Networks}
\newacronym{gnb}{gNB}{Next Generation Node B}
\newacronym{pusch}{PUSCH}{Physical Uplink Shared Channel}
\newacronym{sdr}{SDR}{Software Defined Radio}
\newacronym{phy}{PHY}{Physical Layer}
\newacronym{cu}{CU}{Centralized Unit}
\newacronym{du}{DU}{Distributed Unit}
\newacronym{ru}{RU}{Remote Unit}
\newacronym{ue}{UE}{User Equipment}
\newacronym{ofdm}{OFDM}{Orthogonal Frequency Division Multiplexing}
\newacronym{ofdma}{OFDMA}{Orthogonal Frequency Division Multiple Access}
\newacronym{bf}{BF}{Beam Forming}
\newacronym{mimo}{MIMO}{Multiple-Input, Multiple-Output}
\newacronym{che}{CHE}{Channel Estimation}
\newacronym{dmrs}{DMRS}{Demodulation Reference Symbol}
\newacronym{tti}{TTI}{Transition Time Interval}

\newacronym{add}{add}{Add}
\newacronym{mul}{mul}{Multiply}
\newacronym{mac}{MAC}{Multiply Accumulate}
\newacronym{pmac}{p.mac}{Post-increment Multiply-accumulate}
\newacronym{axpy}{AXPY}{A Times X Plus Y}
\newacronym{dotp}{DOTP}{Dot Product}
\newacronym{sdotp}{SDOTP}{Sum Dot Product}
\newacronym{matmul}{MatMul}{Matrix Multiplication}
\newacronym{gemm}{GEMM}{General Matrix Multiplication}
\newacronym{mvm}{MVM}{Matrix-Vector Multiplication}
\newacronym{fft}{FFT}{Fast Fourier Transform}
\newacronym{sysinv}{SysInv}{Linear System Inversion}
\newacronym{choldec}{CholDec}{Cholesky Decomposition}
\newacronym{mmse}{MMSE}{Minimum Mean Squared Error}
\newacronym{conv2D}{Conv2D}{2D-Convolution}
\newacronym{dct}{DCT}{Direct Cosine Transform}

\newacronym{sram}{SRAM}{Static Random-Access Memory}
\newacronym{dram}{DRAM}{Dynamic Random-Access Memory}
\newacronym{spm}{SPM}{Scratchpad Memory}
\newacronym{tcdm}{TCDM}{Tightly Coupled Data Memory}
\newacronym{IDol}{I\$}{Instruction Cache}
\newacronym{dma}{DMA}{Direct Memory Access}
\newacronym{axi}{AXI}{Advanced eXtensible Interface}
\newacronym{noc}{NoC}{Nework on Chip}
\newacronym{csr}{CSR}{Control Status Register}
\newacronym{hbm}{HBM2E}{High Bandwidth Memory}

\newacronym{ipc}{IPC}{instructions-per-cycle}
\newacronym{wfi}{WFI}{wait-for-interrupt}
\newacronym{raw}{RAW}{read-after-write}
\newacronym{ins}{INS}{instruction}

\newacronym{fpu}{FPU}{Floating Point Unit}
\newacronym{fpss}{FP-SS}{Floating Point Sub-System}
\newacronym{ipu}{IPU}{Integer Processing Unit}
\newacronym{divsqrt}{DIVSQRT}{Division and Square-Root Unit}
\newacronym{lsu}{LSU}{Load Store Unit}
\newacronym{dsp}{DSP}{Digital Signal Processing}

\newacronym{eda}{EDA}{Electronic Design Automation}
\newacronym{ge}{GE}{Gate Equivalent}
\newacronym{fo4}{FO4}{Fan-Out-of-4}
\newacronym{beol}{BEOL}{Back-End-of-Line}
\newacronym{pnr}{PnR}{Place and Route}
\newacronym{ppa}{PPA}{Power, Performance and Area}

\newacronym{numa}{NUMA}{Non-Uniform Memory Access}
\newacronym{fc}{FC}{Fully-Connected}
\newacronym{isa}{ISA}{Instruction Set Architecture}
\newacronym{simd}{SIMD}{Single Instruction Multiple Data}
\newacronym{spmd}{SPMD}{Single Program Multiple Data}
\newacronym{cdf}{CDF}{Cumulative Distribution Function}
\newacronym{api}{API}{Application Programmable Interface}
\newacronym{rtl}{RTL}{Register Transfer Level}
\newacronym{sfr}{SFR}{Synchronization Free Region}
\newacronym{dsl}{DSL}{Domain-Specific Language}

\begin{document}

\ifx\blind\undefined
    \title{A 1024 RV-Cores Shared-L1 Cluster with High Bandwidth Memory Link for Low-Latency 6G-SDR}
    \author{%
      \IEEEauthorblockN{%
        \parbox{\linewidth}{\centering
          Yichao Zhang\IEEEauthorrefmark{1},
          Marco Bertuletti\IEEEauthorrefmark{1},
          Chi Zhang\IEEEauthorrefmark{1},
          Samuel Riedel\IEEEauthorrefmark{1},
          Alessandro Vanelli-Coralli\IEEEauthorrefmark{1}\IEEEauthorrefmark{2} and
          Luca Benini\IEEEauthorrefmark{1}\IEEEauthorrefmark{2}%
        }%
      }%
      \IEEEauthorblockA{%
        \IEEEauthorrefmark{1}ETH Z\"{u}rich, Z\"{u}rich, Switzerland
        \IEEEauthorrefmark{2}Universit\`a di Bologna, Bologna, Italy \\
        Email: 
        yiczhang,
        mbertuletti,
        chizhang,
        sriedel,
        avanelli,
        lbenini,
        @iis.ee.ethz.ch
      }%
      \vspace{-.7cm}
    }
\else
    \title{A 1024 RV-Cores Shared-L1 Cluster with High Bandwidth Memory Link for Low-Latency 6G-SDR}
    \author{\centering{\textit{Authors omitted for blind review.}\vspace{.5cm}}}
\fi
\maketitle

\begin{abstract}
We introduce an open-source architecture for next-generation Radio-Access Network baseband processing: \num{1024} latency-tolerant 32-bit RISC-V cores share \qty{4}{\mebi\byte} of L1 memory via an ultra-low latency interconnect (\num{7}-\num{11} cycles), a modular Direct Memory Access engine provides an efficient link to a high bandwidth memory, such as HBM2E (\SI{98}{\percent} peak bandwidth at \SI{910}{GBps}).
The system achieves leading-edge energy efficiency at sub-\si{\milli\second} latency in key 6G baseband processing kernels: Fast Fourier Transform (\qty{93}{\giga\ops\per\watt}), Beamforming (\qty{125}{\giga\ops\per\watt}), Channel Estimation (\qty{96}{\giga\ops\per\watt}), and Linear System Inversion (\qty{61}{\giga\ops\per\watt}), with only \SI{9}{\percent} data movement overhead.
\end{abstract}


\IEEEpeerreviewmaketitle

\section{Introduction}
Beyond 5G, baseband compute demands rapidly increase with the number of antennas and sub-carriers. For example, \gls{pusch} processes massive-\gls{mimo} transmissions over \SI{100}{\mega\hertz} frequency bandwidth within \SI{1}{\milli\second}.
Energy-efficient many-core \glspl{soc} can provide the flexibility needed to keep up with evolving standards, following the \gls{sdr} paradigm.
However, it is essential for these programmable \glspl{soc} to meet the tight performance and efficiency requirements.

We introduce the \terapool{} \gls{soc}\footnote{\ifx\blind\undefined https://github.com/pulp-platform/mempool \else Open-source information omitted for blind review. \fi}, featuring \num{1024} RISC-V cores, \SI{4}{\mebi\byte} of multi-banked (\num{4096} banks) L1 \gls{spm}, and a minimal overhead link to \gls{hbm} main memory.
Our contributions are:
\textbf{\textcircled{1}} The physical design of a cluster with the largest core count reported for 6G-\gls{sdr} workloads (Tab.~\ref{tab:soa}). Its ultra-low latency core-to-L1 interconnect, operates at up to \qty{924}{\mega\hertz} (TT/\SI{0.80}{\volt}/\SI{25}{\celsius}) in GlobalFoundries' \SI{12}{\nm} FinFET technology.
\textbf{\textcircled{2}} A modular \gls{dma} engine enables transfers between L1 and main memory via a hierarchical \gls{axi} tree.
The open-sourced cycle-accurate \textit{DRAMsys5.0} simulator~\cite{DRAMsys_2022} is used for fast runtime main memory co-simulation\footnote{A QuestaSim linkable dynamic library is created to speed up runtime.}.
We use an \gls{hbm} main memory model to demonstrate our \gls{dma} engine capability to manage ultra-high bandwidth data flows, while the large L1 hides \SI{130}{cycles} average transfer latency, without affecting kernel performance.
We achieve \qty{0.18} to \qty{0.84}{\tera\ops} at \qty{60} to \qty{125}{\giga\ops\per\watt} on key 6G-\gls{sdr} kernels with only \SI{9}{\percent} data movement overhead, while meeting \SI{1}{\milli\second} \gls{pusch} latency requirements at less than \SI{8.8}{\watt} cluster power consumption.

\section{Architecture}
\terapool{}'s \glspl{pe} are single-stage, latency-tolerant \emph{Snitch} cores~\cite{Mempool_2023}.
The cores-L1 interconnect (snapshot Fig.~\ref{fig:cluster_layout}, implementation details Table~\ref{tab:soa}) consists of a 3-level hierarchy of \gls{fc} crossbars (Tile, SubGroup, Group) with pipeline cuts for timing closure (Fig.~\ref{fig:arch_cluster}).
Therefore, access latency varies across the hierarchy and is parametrizable for the point-to-point connections between Groups (1-3-5-7/9/11 cycles) depending on timing constraints.
At the system level (Fig.~\ref{fig:arch_axidma}), we design a modular \gls{dma} split in three: frontend (configuration), midend (transfer split), and backend (data mover).
One Subgroup Tile-shared master \gls{axi} port, supports L1 \gls{IDol} refill or \gls{dma}-controlled data transfers.
The Cluster \gls{axi} masters demultiplex to \gls{dma}-frontend, L2, and \gls{csr}/Peripherals.
Fig.~\ref{fig:cluster_layout} shows ultra-low (\SI{14.1}{\percent}) Cluster area overhead in \gls{ge} for the interconnects (\gls{fc}-L1-crossbar, \gls{axi} \& \gls{dma}).
We connect the L2-\gls{axi} master with two \SI{16}{\gibi\byte} \gls{hbm} stacks.
An address scrambler aligns the data interleaved across \gls{hbm} channels to burst length, reducing \gls{axi} conflicts during transfers.

\section{6G-SDR Key Kernels and Results}
\begin{figure}[ph!]
\centering
\begin{minipage}{\textwidth}
\centering
\includegraphics[width=\linewidth]{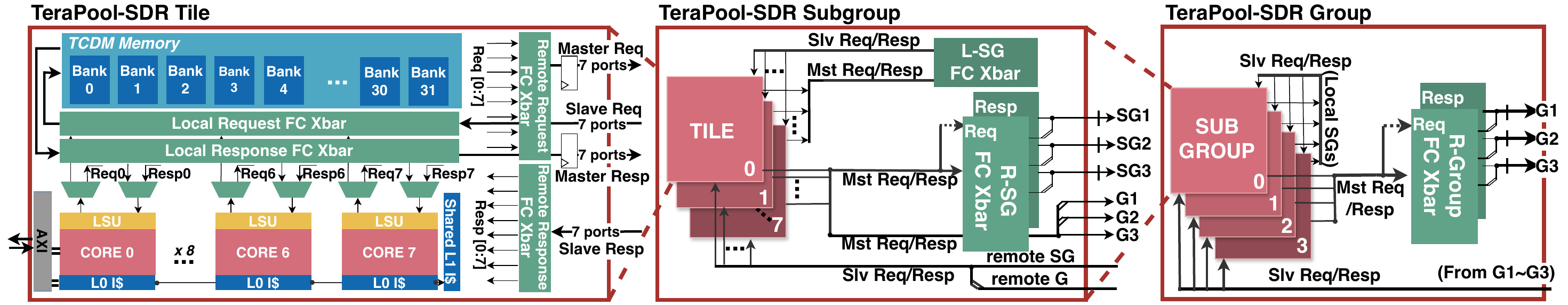}
\vspace{-.6cm}
\caption{Architecture of \terapool{\text{1-3-5-X}}. Subscript stands for cores access latency to banks in each hierarchical level.}
\label{fig:arch_cluster}
\end{minipage}

\begin{minipage}{.49\textwidth}
  \vspace{-4cm}
  \centering
  \includegraphics[width=.9\linewidth]{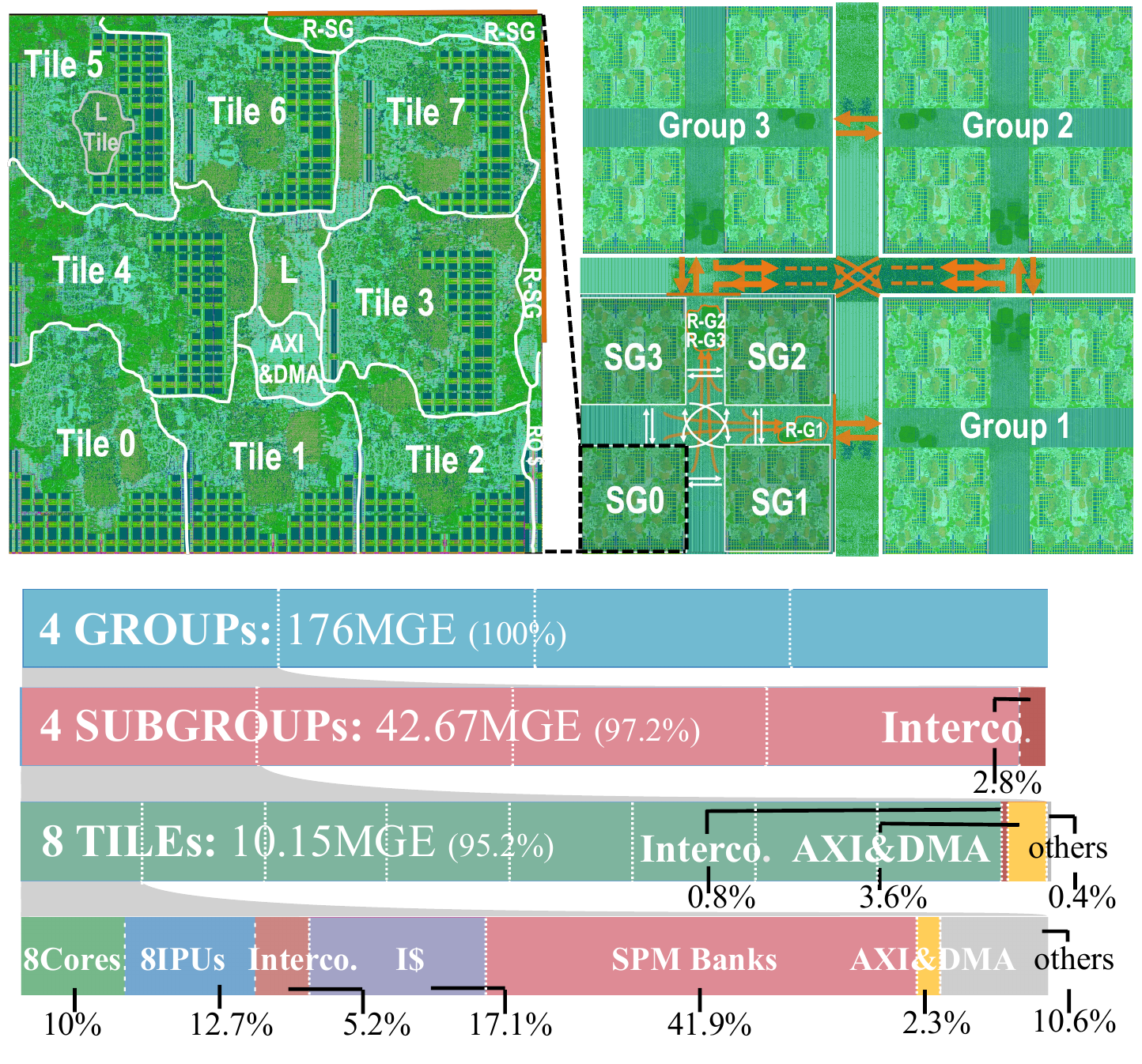}
  \vspace{-.2cm}
  \captionof{figure}{PnR layout view in GlobalFoundries' \SI{12}{\nm} FinFET and hierarchical area breakdown of \terapool{} Cluster.}
  \label{fig:cluster_layout}
\end{minipage}%
\quad
\begin{minipage}{.49\textwidth}
\begin{minipage}{\textwidth}
  \vspace{.1cm}
  \centering
  \includegraphics[width=.9\linewidth]{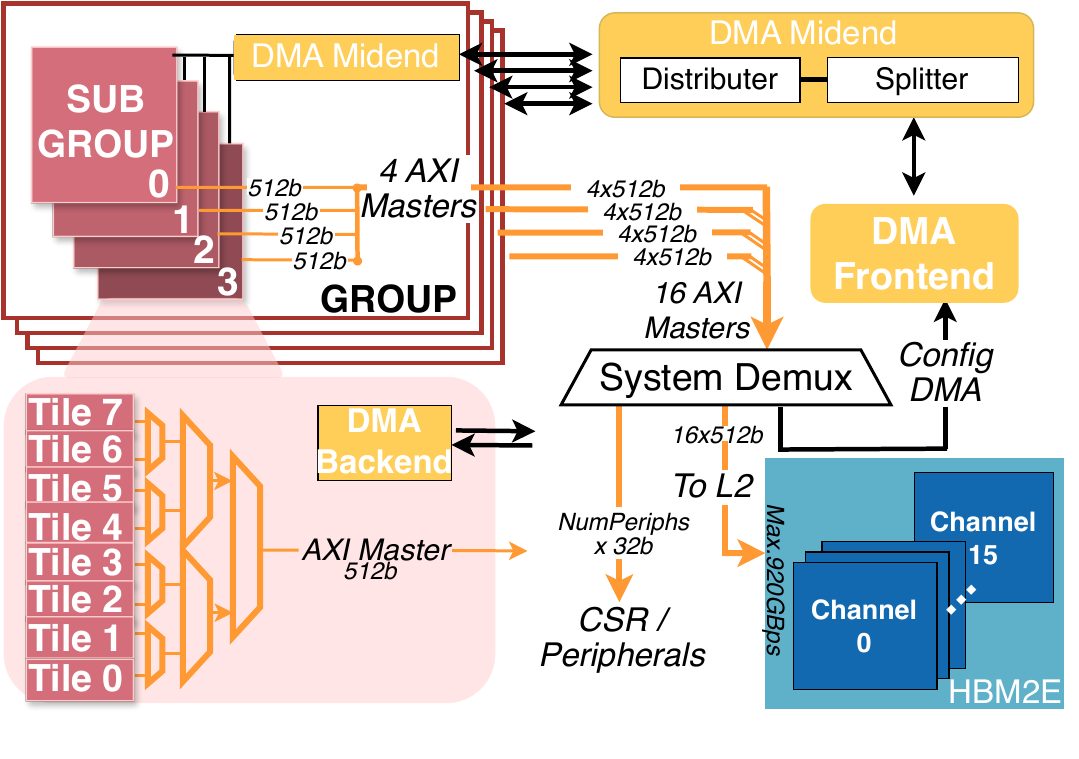}
  \vspace{-.4cm}
  \captionof{figure}{System-level hierarchical AXI interconnection and modular DMA implementation.}
  \label{fig:arch_axidma}
\end{minipage}
\begin{minipage}{\textwidth}
  \vspace{.2cm}
  \centering
  \includegraphics[width=0.9\linewidth]{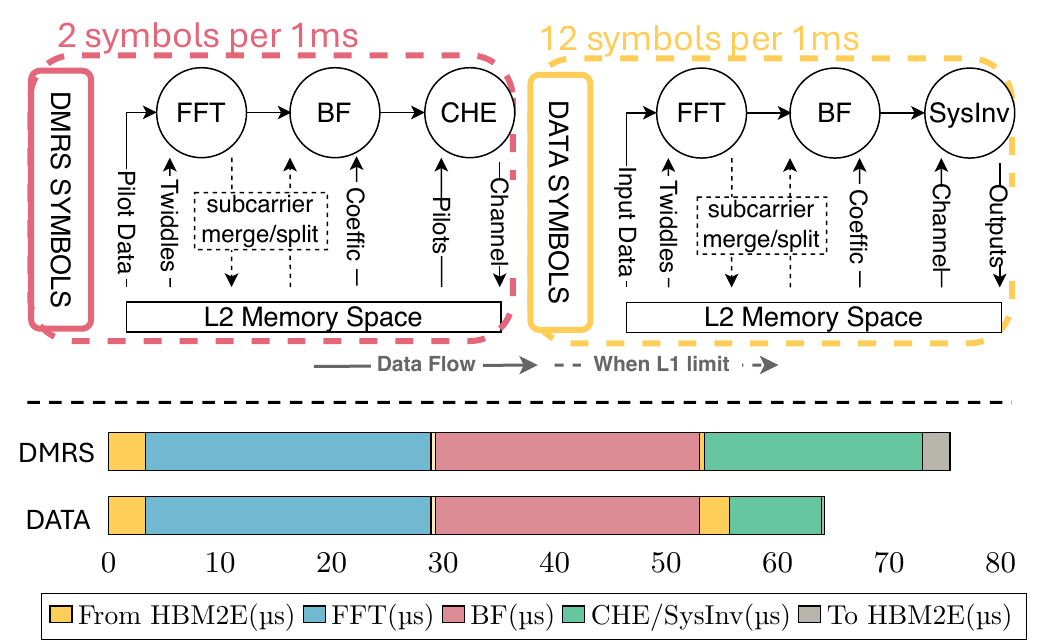}
  \vspace{-.1cm}
  \captionof{figure}{Up: Key kernels' data flow of baseband receiving; Down: Data move vs. compute time for kernels in \terapool{\text{1-3-5-9}}.}
  \label{fig:transfer_overhead}
\end{minipage}
\end{minipage}

\hspace{-.5cm}
\centering
\begin{minipage}{.46\textwidth}
\begin{minipage}{\textwidth}
  \vspace{-4cm}
  \centering
  \includegraphics[width=\linewidth]{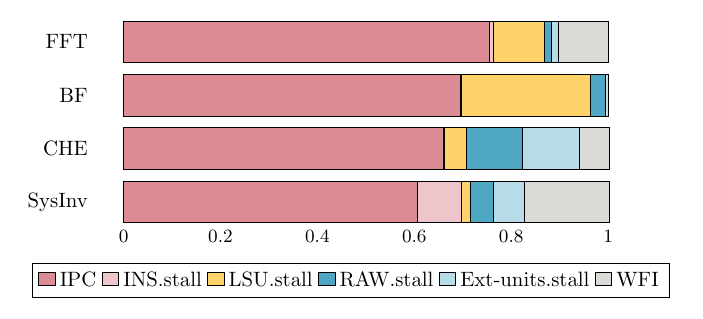}
  \vspace{-.7cm}
  \captionof{figure}{Fraction of instructions and stalls over the total cycles for the kernel's execution in \terapool{\text{1-3-5-9}}.}
  \label{fig:ipc_sdr}
\end{minipage}%

\begin{minipage}{\textwidth}
  \vspace{-.1cm}
  \centering
  \includegraphics[width=.9\linewidth]{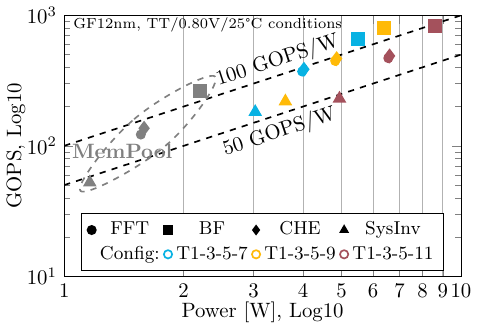}
  \vspace{-.2cm}
  \captionof{figure}{\terapool{} Energy Efficiency for SDR workloads, compared with a $4
  \times$ smaller \textit{MemPool} cluster.}
  \label{fig:energy_sdr}
\end{minipage}%
\section*{Acknowledgment}
\ifx\blind\undefined
    Funded in part by COREnext supported by Horizon Europe program under grant agreement No. \num{101092598}.
\else
    Information omitted for blind review.
\fi
\end{minipage}
\quad
\begin{minipage}{0.49\textwidth}
\hspace{.5cm}
\begin{minipage}{\textwidth}
  \vspace{.2cm}
  \centering
  \captionof{table}{State of Art Comparison}
  \resizebox{\textwidth}{!}{%
    \begin{tabular}{@{}c|c|ccc@{}}
    \toprule
    & \begin{tabular}[c]{@{}c@{}}This Work\\ GF12\end{tabular}   & \begin{tabular}[c]{@{}c@{}}NVIDIA H100\\ TSMC4N~\cite{Nvidia_2023}\end{tabular} & \begin{tabular}[c]{@{}c@{}}Kalray MPPA3-80\\ TSMC16~\cite{kalary_2022}\end{tabular} & \begin{tabular}[c]{@{}c@{}}Ramon RC64\\ TSMC65~\cite{Ramon_2021}\end{tabular} \\ \midrule
    Frequency (MHz)       & (730-924)@(7-11)cycles       & 1755         & 600$\sim$1200      & 300  \\
    Cores / Shared L1     & 1024 / 4MiB         & 128 / 256KB    & 16 / 4MB  & 64 / 4MB     \\
    \begin{tabular}[c]{@{}c@{}}Shared L1 latency cycles\\ L1 Throughput\end{tabular} & \begin{tabular}[c]{@{}c@{}}1$\sim$7/9/11 \\ 4KiB/Cyc\end{tabular} & \begin{tabular}[c]{@{}c@{}}16.57\\ 0.125KiB/Cyc\end{tabular} & \begin{tabular}[c]{@{}c@{}}N.A.\\ 0.25KiB/Cyc\end{tabular}       & \begin{tabular}[c]{@{}c@{}}N.A.\\N.A.\end{tabular}                                                        \\
    \begin{tabular}[c]{@{}c@{}}Main Mem Latency\\ Main Mem Throughput\end{tabular}         & \begin{tabular}[c]{@{}c@{}}130ns\\ 920GBPS\end{tabular}     & \begin{tabular}[c]{@{}c@{}}377.9ns\\ 2TBPS\end{tabular}      & \begin{tabular}[c]{@{}c@{}}N.A.\\ 600GBPS\end{tabular}           & \begin{tabular}[c]{@{}c@{}}N.A.\\ 0.7GBPS\end{tabular}      \\
    Power / Shared-L1 Cluster  & \SI{8.8}{\watt}     & \SI{6.14}{\watt}   & \SI{4.9}{\watt}  & \SI{5}{\watt}\\
    Energy-Eff (INT32)  & 125GOPS/W     & 73.15GOPS/W   & N.A.  & 15GOPS/W 
    \end{tabular}}
    \label{tab:soa}
\end{minipage}%

\begin{adjustwidth}{0.5cm}{0cm}
\bibliographystyle{IEEEtran} 
\bibliography{bibliography.bib}
\end{adjustwidth}
\end{minipage}
\end{figure}

Fig.~\ref{fig:transfer_overhead} shows key kernels flow for \gls{pusch} \glspl{dmrs} and Data symbols.
We consider \num{64} antennas, \num{3276} sub-carriers, \num{32} beams, \num{4} transmitters~\cite{PUSCH_2023}.
The problem size is large.
For example, in small clusters, when \gls{fft} data does not fit in L1, demodulated sub-carriers must be stored in L2 before \gls{bf}. Multiplication by \gls{bf} coefficients is tiled.
\terapool{} with its workload-sized \SI{4}{\mebi\byte} shared L1, reduces the required data splitting significantly.
Data transfer overhead through \gls{hbm} (\SI{910}{GBps}, \SI{98}{\percent} efficiency) for all benchmarked kernels is $<$\SI{9}{\percent} (Fig.~\ref{fig:transfer_overhead}), even in presence of a considerable \SI{130}{cycles} average latency.
All kernels have $>$\num{0.6} \gls{ipc} (Fig.~\ref{fig:ipc_sdr}), as we exploit data locality and proximity of memory hierarchies to minimize \gls{lsu} stalls from contentions to the interconnect shared resources.
\terapool{} achieves energy efficiency greater or equal than a \num{4}$\times$ smaller, \num{256} cores shared-\SI{1}{\mebi\byte} MemPool cluster~\cite{Mempool_2023}, thanks to its low-power interconnect (Fig.~\ref{fig:energy_sdr}).
\terapool{\text{1-3-5-9}} is optimal for energy efficiency, and \terapool{\text{1-3-5-11}} excels in performance for the selected \gls{sdr} workload.
Achieved $<$\SI{70}{\micro\second} latency per data symbol at a power consumption below \SI{8.8}{\watt} demonstrates that \terapool{} is suitable for baseband operations.

\onecolumn

\end{document}